\documentclass [11pt,letterpaper]{JHEP3}
\usepackage{epsfig,array,cite,latexsym,amsmath,oldgerm}
\usepackage{graphicx}
\usepackage{epsfig}
\usepackage{epic}       
\usepackage{cite}	
\usepackage{psfrag}     
\usepackage{bm}		
\usepackage{verbatim}   
\advance\parskip 1.0pt plus 1.0pt minus 2.0pt	
\def\coeff#1#2{{\textstyle {\frac {#1}{#2}}}}
\def\half{\coeff 12}

\def\tr{{\rm tr}}

\def\beq{\begin{equation}}
\def\eeq{\end{equation}}
\newcommand{\drawsquare}[2]{\hbox{%
\rule{#2pt}{#1pt}\hskip-#2pt
\rule{#1pt}{#2pt}\hskip-#1pt
\rule[#1pt]{#1pt}{#2pt}}\rule[#1pt]{#2pt}{#2pt}\hskip-#2pt
\rule{#2pt}{#1pt}}
\newcommand{\Yfund}{\raisebox{-.5pt}{\drawsquare{6.5}{0.4}}}
\newcommand\bal{ \begin{align}}
\newcommand\eal{\end{align} }

\newcommand\eqn[1]{\label{eq:#1}} 
\newcommand\Eq[1]{Eq.~\eqref{eq:#1}} 

\newcommand\bfZ{\mathbf{Z}}
\newcommand\bfl{\boldsymbol{\Lambda}}
\newcommand\bfXi{\boldsymbol{\Xi}}

\newcommand{\CN}{{\cal N}}

\newcommand{\CQ}{{\cal Q}}

\newcommand{\bfn}{{\bf n}}

\newcommand{\ah}{\mathbf{\hat{e}}_a}

\DeclareMathOperator{\Tr}{Tr\,}

\newcommand{\sla}[1]%
        {\kern .25em\raise.18ex\hbox{$/$}\kern-.55em #1}

\newcommand{\mybar}[1]%
        {\kern 0.6pt\overline{\kern -0.6pt#1\kern -0.6pt}\kern 0.6pt}
\newcommand{\dig}{\kern-1.5pt \raisebox{.9ex}{$\cdot$}  \kern1.5pt
  \raisebox{0ex}{${\mathbf\cdot}$}\kern1.5pt \raisebox{-.9ex}{$\cdot$}} 
\newcommand{\digb}{\kern-1.5pt \raisebox{.75ex}{$\cdot$}  \kern1.5pt
  \raisebox{0ex}{${\mathbf\cdot}$}\kern1.5pt \raisebox{-.75ex}{$\cdot$}} 
\newcommand{\digc}{\kern-1.5pt \raisebox{1.05ex}{$\cdot$}  \kern1.5pt
  \raisebox{0ex}{${\mathbf\cdot}$}\kern1.5pt \raisebox{-1.05ex}{$\cdot$}} 
\newsavebox{\flippedL}
\savebox{\flippedL}
    {
    \setlength{\unitlength}{0.7mm}
    \begin{picture}(1,3) (3,0.5)
	    \drawline(0,0)(4,0)
            \drawline(4,0)(4,4)
	    \drawline(4,4)(3,4)
	    \drawline(3,4)(3,1)
	    \drawline(3,1)(0,1)
            \drawline(0,1)(0,0)
        \end{picture}
    }
  \preprint{BUHEP 05-07}
\title
    {%
Compact Gauge Fields  for  Supersymmetric Lattices
    }%
\author
    {%
     Mithat  \"Unsal 
    \\Department of Physics,
    Boston University,  Boston, MA 02215
    \\Email: 
    \parbox[t]{2in}{\email {unsal@buphy.bu.edu} }
    }
\abstract
    {%
 We show that a large class of Euclidean 
extended supersymmetric lattice gauge theories  
constructed  in \cite{Cohen:2003aa,Cohen:2003qw, Kaplan:2005ta} can be 
regarded as  compact formulations 
by using the  polar  decomposition of the  complex link fields. 
In particular, the gauge
part of the supersymmetric lattice action is the standard  Wilson
action. This  formulation facilitates the construction of gauge invariant 
operators.
  }%
\keywords{Supersymmetry, Lattice Gauge Field Theories}

\begin{document}
\setlength{\baselineskip}{1.25\baselineskip}

\section{Introduction and results}
\label{sec:intro} 
In \cite{Kaplan:2002wv, Cohen:2003aa, Cohen:2003qw, Kaplan:2005ta}, the spatial and 
Euclidean spacetime lattices 
whose continuum limits are extended supersymmetric gauge theories in various
dimensions are  constructed.   
These lattices are obtained by the orbifold projection of  supersymmetric 
matrix models.  The projection generates a lattice, 
while preserving a subset
of supersymmetry. But, a priori, a dimensionful parameter 
  which can be identified  as a lattice spacing is absent.  However, these
  theories  possess a moduli space of vacua, along which the potential
  vanishes, and the distance from the origin of the moduli space is 
interpreted as an inverse lattice spacing.  
Moving to infinity  in the 
 moduli  space corresponds to taking the lattice spacing to zero. The
 construction of the supersymmetric lattices  is inspired by orbifold 
projection  \cite{Douglas:1996sw}
and  the deconstruction of the supersymmetric theories 
\cite{ArkaniHamed:2001ca, ArkaniHamed:2001ie, Rothstein:2001tu}.

In   \cite{Kaplan:2002wv, Cohen:2003aa, Cohen:2003qw, Kaplan:2005ta }, 
the fluctuations of the  complex  bosonic link fields around a particular 
configuration in moduli space  
are split into an
hermitian and antihermitian   
matrix, as in
the cartesian decomposition of a complex number. In the
continuum limit,  these matrices  give rise to scalars and the 
gauge bosons of the target theory respectively. This construction  provides a  
formulation in which the   gauge boson is noncompact. However, this expansion 
is not the most natural one for two reasons. In terms of the shifted
variables, certain global symmetries and the local gauge symmetry 
become hidden. Let us explain both in turn. The orbifold matrix theory, 
i.e, the lattice action before the shift, possesses  at least a 
 $U(1)^d$ symmetry 
\footnote{There are other continuous  symmetries of the orbifold matrix
  theory, for example, the R-symmetry.}    
which corresponds to rotations of the complex link fields as 
$z_{a,\bfn} \rightarrow  e^{i\alpha_{b}\delta_{ab}}  z_{a,\bfn}$.  
In terms of shifted variables, this symmetry is not manifest in the action.
The latter issue is gauge covariance. The complex link  boson  
$z_{a,\bfn}$ is in bifundamental representation and transforms covariantly
under gauge transformation. Consequently,   
the orbifold matrix theory  action is manifestly gauge invariant. 
However, in terms of shifted fields, the gauge invariance is a nonlinearly 
realized symmetry. 
It is, therefore, desirable, to make both the global symmetry and gauge
symmetry manifest.    
   
In this note, we  use a parametrization of the complex link fields
which preserves the exact gauge invariance of the orbifold matrix theory 
and keeps  global symmetries  manifest. It
relies on  the polar decomposition of a complex matrix  into radial 
and angular variables. We expand our action 
around the same point  in the  moduli space.  
  This  expansion  generates a lattice  action which is  manifestly gauge
  invariant, and the gauge fields are compact, unitary  matrices, hence there 
is no need for  gauge-fixing.   However, having a compact gauge integration 
is  not one of the advantages, since the moduli spaces of our target 
theories are  noncompact.   On the other hand, 
the polar decomposition  preserves the exact global symmetries of 
the orbifold matrix theory and the observables of the theory are 
charged under these symmetries. Such observables are the order parameters 
for the corresponding symmetries. 
\footnote{
 We  want to point out that our construction is not a new 
lattice formulation of the supersymmetric lattice gauge theories; it is 
merely a  reparametrization of the complex link field.}

To facilitate the identification  of the gauge part of our supersymmetric
lattice action with  the standard Wilson action  \cite{Wilson:1974sk},  
we choose to work with hypercubic lattices  on the moduli space. 
As 
explained in detail in \cite{Kaplan:2002wv, Kaplan:2005ta}, 
different points in moduli space 
correspond to different structures of the unit cell on the lattice. For
example, in  the two dimensional lattice for the $\CQ=16$ supercharge 
 target theory,   
 different points in moduli space
correspond to a hexagonal lattice, square  lattice with a diagonal or 
asymmetric lattices such as a rectangular lattice with diagonal. Each of
these lattices has a different point group symmetry. There are trajectories 
in moduli space which respect a particular type of point group symmetry. To
take the continuum limit, one moves along a particular trajectory out to
infinity with an appropriate scaling of the number of sites on the lattice. 
We chose  previously  \cite{Kaplan:2002wv} the most symmetric lattices to
minimize the number of relevant and marginal operators. Our purpose  here is 
to show a correspondence with the Wilson action, and the hypercubic 
trajectories are better suited for that.  
\setlength{\extrarowheight}{0pt}
\begin{table}[t]
\centerline{
\begin{tabular}{|c|c|c|c|}
\hline
 & $\CQ=4$ & $\CQ=8$ & $\CQ=16$
\\ \hline
$ d=1 $& $ 2  $&$ 4 $&$ 8 $
\\\hline
$ d=2 $& $ 1  $&$ 2 $&$ 4 $
\\
\hline
$ d=3 $&{\Large $\times$}   &$ 1 $&$ 2 $
\\ \hline
$ d=4 $& {\Large $\times$}  & {\Large $\times$}  &$ 1 $
\\ \hline
\end{tabular}
}
\caption{\sl The 
Euclidean lattices for extended SYM theories. $\CQ$ is the 
amount of supersymmetry in both the mother theory and the target theory. 
The numbers in the boxes 
corresponds to the number of exactly realized supersymmetries on the 
Euclidean lattice.   The  boxes with  {\Large $\times$} are the theories that 
could not be reached due to insufficiency of R-symmetry within our approach. 
\label{tab:tab1}}
\end{table}

The approach that has been presented here is applicable to all of the 
Euclidean spacetime lattices listed in Table \ref{tab:tab1}, in particular, 
 including all the $ \CQ=16$ target theories in $d \leq 4$ dimensions.
\footnote{These theories are sometimes  named with respect to 
 the multiplicity of their  minimal
spinor dimensions in the corresponding dimension. More conventional names for
the $\CQ=4, \CQ=8, \CQ=16$  theories are: in $d=4$ dimensions $\CN=1,2,4$,  
in $d=3$ dimensions $\CN=2,4,8$, and  in $d=2$ dimensions $\CN=(2,2), (4,4),
(8,8)$ respectively. 
 }
In  Table  
\ref{tab:tab1}, each  box is  associated with a pair $(d, \CQ)$, a $d$
dimensional  target SYM theory with $\CQ$ supersymmetries. 
The number within the box (associated with  $(d, \CQ)$ theory) 
corresponds to the number of exactly 
realized supersymmetries by the $d$ dimensional lattice action.
Among these, there is also the $\CN=4$ SYM theory in 
$d=4$ dimensions \cite{Kaplan:2005ta}. The theories with  
 {\Large $\times$} sign  
are the ones that we are unable to reach,  within our approach, due to 
insufficiency of the R-symmetry of the mother theory. 
The details of each supersymmetric lattice gauge theory can be quite 
different. For example, the number of exactly realized supersymmetries, 
the supersymmetry algebra, 
the multiplet structures,  the point group symmetries of the  
spacetime lattices can be different in each case. 
The application in section \ref{sec:appl-comp-latt}  is aimed to show the
realization of the ideas in one simple example, the $\CQ=4$ target theory in
$d=2$ dimensions. Generalization to other 
supersymmetric lattice theories listed in  Table \ref{tab:tab1} is also 
possible. 
 
Before showing the realizations of these ideas explicitly
 in a simple
example, we first want to point out  the other 
 recent approach  to  the supersymmetic lattice, which is sometimes 
known under the rubric twisted supersymmetry.  
(For references to earlier work, see  
\cite{Kaplan:2002wv, Kaplan:2003uh,Feo:2002yi}).
Catterall and collaborators implemented the supersymmetry on the
lattice by keeping one or more 
nilpotent supersymmetries exactly realized on the 
lattice \cite{Catterall:2001fr, Catterall:2003uf}.  This led to the 
construction of the two dimensional 
Wess-Zumino and supersymmetric  sigma models in various dimensions. 
\cite{Catterall:2003ae, Ghadab:2004rt }
In \cite{Giedt:2004qs},  the general criteria under which a 
nilpotent symmetry can be carried consistently with  a latticization 
had been analyzed(also see \cite{D'Adda:2004jb, D'Adda:2004ia}, which 
advocates more, and \cite{Bonini:2005qx} for the recent results on $N=1$
Wess-Zumino in $d=4$).  
Recently, Sugino \cite{Sugino:2003yb,Sugino:2004qd,Sugino:2004uv}  
generalized this approach to the gauge theories with compact gauge fields 
in the formulation. 
He showed 
that  a continuum supersymmetric gauge theory written in a $Q$-exact form, can
be carried to the lattice and gave the lattice counterpart of
$Q$-transformations. 
However, this procedure is not  unique.
And as a result,  he 
 encounters a vacuum  degeneracy problem on the gauge field sector, as well
 as spurious zero modes in the theory. Sugino argues that both  problems 
can be solved. (see \cite{Sugino:2004qd, Sugino:2004uv} for details.) 
More recently, Catterall \cite{Catterall:2004np} 
constructed the two  dimensional four supercharge  SYM 
theory by using a geometric approach 
that is free of the problems encountered in  \cite{Sugino:2003yb}. (Also see 
  \cite{Catterall:2005fd}). However, the lattice theory he constructs
  requires a complexification (hence doubling) of all the degrees of
  freedom. This turns,   for example, the unitary compact variables, into 
noncompact variables.  Then, he
  conjectures that, one  can  restrict the path integral   to a real line 
in the  field space by preserving the Ward identities associated with
twisted supersymmetry, and also recovers the desired target theory. The
approach that we will follow  does not encounter this problem.    

\section{Application: Compact lattice action  
for $\CN=(2,2)$ SYM theory }
\label{sec:appl-comp-latt}
To illustrate these ideas explicitly, we work through a simple example  in
detail. The target theory is the  $\CN=(2,2)$ SYM theory in $d=2$
dimensions. The lattice construction is examined in detail  in  
\cite{Cohen:2003aa}, and we refer the reader there for more
details. 
The arguments in this section can be easily applied to any of
the other target theories listed in Table \ref{tab:tab1}.  
 
The lattice action
for the  $\CN=(2,2)$  SYM target theory with gauge group 
$U(k)$,  
in $\CQ=1$ 
superfield formulation is  given by \cite{Cohen:2003aa}  
\begin{equation}
\begin{aligned}
S = \frac{1}{g^2} \sum_{\bfn} \tr\int d\theta\,\Bigl[& 
-\half \bfl_\bfn 
 \partial_\theta  \bfl_\bfn -  \bfl_\bfn  \left(\mybar
    z_{a,\bfn-\ah}  
    \bfZ_{a,\bfn-\ah} - \bfZ_{a,\bfn}  \mybar
    z_{a,\bfn}\right) \\
&
-  \bfXi_{12, \bfn}  ( \bfZ_{1,\bfn}  \bfZ_{2,\bfn
  +{\bf \hat e_1}} - \bfZ_{2,\bfn}  \bfZ_{1,\bfn + {\bf \hat e_2} })
\Bigr]
\eqn{ssact2} \,,
\end{aligned}
\end{equation} 
where $\bfn$ is a two component integer vector labeling sites on the lattice 
and ${\bf \hat e_a}$ is the unit vector in a'th direction where  $a=1,2$.
\footnote{The lattice action \Eq{ssact2} can also be expressed in a 
$Q$-exact  
form (also noted in \cite{Giedt:2004tn}), which is given by 
\begin{equation}
\begin{aligned}
S = \frac{1}{g^2} \sum_{\bfn} \tr Q \Bigl[& 
+\half \lambda_\bfn (id_\bfn) -  \lambda_\bfn  \left(\mybar
    z_{a,\bfn-\ah}  
    z_{a,\bfn-\ah} - z_{a,\bfn}  \mybar
    z_{a,\bfn}\right) 
-  \xi_{12,\bfn}  ( z_{1,\bfn}  z_{2,\bfn
  +{\bf \hat e_1}} - z_{2,\bfn}  z_{1,\bfn + {\bf \hat e_2} })
\Bigr]
\eqn{Qexact} \,,
\end{aligned}
\end{equation} 
The action of $Q$ on the components can be read of from  the supermultiplets 
\Eq{superfield1}.
For example, $Qz_{a,\bfn}= \sqrt 2 \psi_{a,\bfn} , \ Q \psi_{a,\bfn}=0$ etc. 
The supersymmetry algebra is $Q^2 \cdot =0$. This is, in the sense of supersymmetry
algebra, the  difference with the Catterall's construction 
\cite{Catterall:2004np}, in which 
the square of the twisted supersymmetry generator  is an infinitesimal field 
dependent gauge rotation, $Q^2 \cdot =\delta_{gauge} \cdot $. 
}
The $\CQ=1$  supersymmetric  multiplets in terms of their component fields 
can be expressed as 
\begin{equation}
\begin{aligned}
{\bf \Lambda}_{\bfn}&= \lambda_{\bfn} -i\theta  d_{\bfn} \ ,\\
%
%
{\bfZ}_{a,\bfn} &= z_{a,\bfn} + \sqrt{2}\,\theta \,\psi_{a,\bfn}\ ,\\
%
%
{\mybar z}_{a,\bfn} &= {\mybar z_{a,\bfn}}, \\
%
%
{\bf \Xi}_{12,\bfn}&= \xi_{12,\bfn} -  2\theta\,\, (\mybar
  z_{2,\bfn+ {\bf \hat e_1} }  \mybar z_{1,\bfn} - 
\,\,
  \mybar z_{1,\bfn+ {\bf \hat e_2} } \mybar z_{2,\bfn})
 \ .
\label{eq:superfield1}
\end{aligned}
\end{equation}
The fermi multiplet ${\bf \Lambda}_{\bfn}$ lives on the site $\bfn$.  
The bosonic multiplet ${\bfZ}_{a,\bfn}$ lives on the oriented link, 
starting at   $\bfn$ and ending at ${ \bfn + \bf \hat e_a}$. The supersymmetry
singlet ${\mybar z}_{a,\bfn}$  lives on the oppositely oriented link, 
starting at   ${\bfn + \bf \hat e_a}$ and ending at $\bfn$, and the 
diagonal 
fermi  multiplet ${\bf \Xi}_{\bfn}$ resides on the diagonal link, starting at
 ${\bfn + \bf \hat e_1 + \hat e_2 }$ and ending at  $\bfn$. The lowest 
 component fermi multiplet ${\bf \Xi}_{12,\bfn}$  
 is antisymmetric under the exchange of its subscripts;
$\xi_{12}=-\xi_{21}$,
as its supersymmetric partner.
By substituting the multiplets into the action \Eq{ssact2},  we obtain 
the action in component fields.  For convenience, we  split the action into 
the bosonic  and fermionic parts. The bosonic part is  
\begin{equation}
\begin{aligned}
S_b = \frac{1}{g^2} \sum_{\bfn} \tr \Bigl[&
\half d_\bfn^2   +  id_\bfn  \left(\mybar
    z_{a,\bfn-\ah}  
    z_{a,\bfn-\ah} - z_{a,\bfn}  \mybar
    z_{a,\bfn}\right) + 2 |( z_{1,\bfn} z_{2,\bfn
  +{\bf \hat e_1}} - z_{2,\bfn}  z_{1,\bfn + {\bf \hat e_2} })|^2 
\Bigr]
\eqn{ssact2b} \,,
\end{aligned}
\end{equation} 
and the fermionic part is 
\begin{equation}
\begin{aligned}
S_f = \frac{\sqrt 2 }{g^2} \sum_{\bfn} \tr \Bigl[
 \lambda_\bfn \left(\mybar
    z_{a,\bfn-\ah}  
    \psi_{a,\bfn-\ah} - \psi_{a,\bfn}  \mybar
    z_{a,\bfn}\right)
&
+ \xi_{ab,\bfn}  ( z_{a,\bfn}  \psi_{b,\bfn
  +{\bf \hat e_a}} - \psi_{b,\bfn}  z_{a,\bfn + {\bf \hat e_b} })
\Bigr] \ .
\eqn{ssact2f} \, 
\end{aligned}
\end{equation} 

The construction,
symmetries and the continuum limit of this lattice action had been examined 
in detail in \cite{Cohen:2003aa}.  The symmetries are the $U(k)$ gauge 
symmetry, the 
discrete translations $Z_N \times Z_N$ of the lattice, a $Z_2$ point
group symmetry, the $\CQ=1$ supersymmetry and a $U(1)^3$ global symmetry
given in Table \ref{tab:tab2}.   The last $U(1)$ denoted as
$Y$ in Table \ref{tab:tab2} is an R-symmetry on the lattice, i.e., it does not 
commute with supersymmetry since the 
the  superspace coordinate $\theta$  is charged under it.  
The global  symmetries  will be an essential part of the discussion 
of observables. 
\setlength{\extrarowheight}{5pt}
\begin{table}[t]
\centerline{
\begin{tabular}
{|c||c|c|c|c||c|c|c|}
\hline
&$  \bfZ_{1,\bfn}  $&$\mybar  z_{1, \bfn} $&$ \bfZ_{2,\bfn}$&$\mybar  z_{2,\bfn}$
&$ \bfl_{\bfn} $&$
{\bf \Xi}_{\bfn} $&$\theta$
\\ \hline
$ r_1 $&$ +1 $&$ -1 $&$ \,\ 0 $&$ \,\ 0 $&$ \,\ 0 $&$ -1 $&$  0 $
\\
$ r_2 $&$\,\ 0 $&$ \,\ 0 $&$ +1 $&$ -1 $&$ \,\ 0 $&$ -1 $&$ \,\ 0 $  
\\
%
$ Y $ &$  \,\ 0 $&$ \,\ 0 $&$ \,\ 0 $&$ \,\ 0 $&$ +\half $&$ +\half
$&$ +\half $
\\ \hline
 \end{tabular}
}
\caption{\sl The  $U(1)^3$ global symmetry charges 
 of the  lattice theory. The  
 $r_{1,2}$ are the ones that are used in orbifold projection. The last one, 
denoted by $Y$, is the exact R-symmetry on the lattice.  
\label{tab:tab2}}
\end{table}

In \cite{Cohen:2003aa}, we expanded the lattice  action  \Eq{ssact2} 
around a point in moduli space, 
where the vacuum expectation value of the link field is interpreted as 
the inverse lattice spacing. Thus, the complex link fields had been written 
as 
\begin{equation}
z_{a,\bfn}= \frac{1}{\sqrt 2 a } {\bf 1}_k +  \frac{h_{a,\bfn} + 
i v_{a,
    \bfn}}{\sqrt 2} 
\eqn{cartesian}
\end{equation}
where $h_{a, \bfn}$ and   $v_{a, \bfn}$ are the Hermitian matrices, which
become the scalar and vector boson of the continuum theory. 
However, in 
terms of shifted fields, the  $U(1)^2$ subgroup  of the  $U(1)^3$, 
associated with $(r_1, r_2)$
charges in Table \ref{tab:tab2}, is hidden. Even though 
these symmetries are there, they become obscured, and we can not benefit 
from them easily. (To better appreciate these symmetries, one should 
address the  observables, see section \ref{sec:observables})   
It is preferable to make these  symmetries manifest. 
The other point is gauge covariance.  The lattice action \Eq{ssact2} is
gauge invariant and its constituents transforms covariantly under gauge 
rotations.. 
However, the shifted fields hide manifest gauge covariance and it is hard 
to construct manifestly gauge invariant objects. 
In this sense, 
both a $U(1)^2$ subset of the global symmetries and gauge symmetry are
nonlinearly realized when the lattice action is expressed in terms of shifted
fields.  It is, therefore,  preferable to make the global symmetries and 
gauge invariance manifest.

Here, we show that there is a more natural decomposition of the 
complex bosonic link fields, 
which makes all the global symmetries manifest and generates manifestly gauge 
invariant hopping terms.  It is  the polar  decomposition of the 
complex link matrices.  This also provides a formulation of the
supersymmetric lattice gauge theory in which the gauge fields are 
compact, group valued matrices.  However, it is not a new formulation, it is 
reexpressing the \Eq{ssact2} in a new parametrization. 

Given  a complex nonsingular matrix $z_{a,\bfn}$,  we can always write  
$z_{a,\bfn}
= H_{a,\bfn}U_{a,\bfn}$ where 
$H_{a,\bfn}$ is a  hermitian 
nonnegative matrix and $U_{a,\bfn}$ is a unitary matrix. 
This decomposition is unique (modulo left-right decomposition, we chose left.)
up to a set of measure zero. It is not unique  for matrices with
zero eigenvalues. However, this is not a problem for us, because we do an
expansion about a point which is far from the origin of the moduli space.  
The vacuum that we are expanding around is    
$ \langle H_{a,\bfn}\rangle = \frac{1}{\sqrt{2}a }1_k$ and $U_{a,\bfn}=1_k$ 
.  Thus, we express  the complex link  matrices as 
\begin{equation}
z_{a,\bfn}=\frac{1}{\sqrt 2} H_{a,\bfn} U_{a,\bfn} =  
\frac{1}{\sqrt 2} \; (\frac{{\bf 1}_k}{a} + h_{a,\bfn}) U_{a,\bfn}
\eqn{polar}
\end{equation}
where  $H_{a,\bfn}$ is a Hermitian matrix   with nonnegative eigenvalues, 
$U_{a,\bfn}$ is a unitary matrix and  $h_{a,\bfn}$ is a Hermitian matrix.  
The  $\sqrt 2$ in the denominator is for  normalization.  
Notice that for the small gauge field configurations, one can expand the 
unitary matrix $U_{a,\bfn}= 1+ i a v_{a, \bfn} + O(a^2)$ and 
the expression  for the polar 
decomposition reduce to cartesian decomposition \Eq{cartesian}. Thus, for
small gauge field configurations, the corresponding lattice action reduces to
the action we examined in  \cite{Cohen:2003aa}. For this reason, we will
not reexamine the classical and the quantum continuum limit and the 
matching of the
target theory fields to the lattice fields. 

Now, let us analyze briefly the gauge transformation  properties.
Notice that the first term in \Eq{polar} dictates the  gauge transformation 
property of the unitary link matrix  $U_{a,\bfn}$. 
Let $g_{\bfn}$ denote a unitary
gauge rotation matrix.  Since under a  gauge
transformation, the complex link matrix transforms as  bifundamental
$z_{a,\bfn} \rightarrow    g_{\bfn}  z_{a,\bfn}  g_{\bfn +  \ah}^{\dagger}$,
the 
unitary link field  transforms the same way as well, 
$U_{a,\bfn} \rightarrow    g_{\bfn}  U_{a,\bfn}  g_{\bfn +  \ah}^{\dagger}$ and the
hermitian matrix  as an adjoint of site $\bfn$, 
$h_{a,\bfn} \rightarrow    g_{\bfn}
h_{a,\bfn}  g_{\bfn }^{\dagger}$. Basically, the polar decomposition places the 
scalar   
$h_{a,\bfn}$ on the site  $\bfn$ and $U_{a,\bfn}$ on the oriented link 
$(\bfn, \bfn+ \ah)$, where  the first entry is the starting point and the 
latter is
the termination point. 

The global $U(1)^2$  symmetry, which becomes  hidden in the case of cartesian
decomposition,  is manifest now. For example,  under the first two $U(1)^2$,  
the variable 
$z_{1,\bfn}$ has a global charge $(1,0)$. This imposes  the charge of 
the unitary link field  $U_{1,\bfn}$ as $(1,0)$,  
the charge of the hermitian scalar $h_{1,\bfn}$ as 
$(0,0)$. The charge of the fermionic superpartner of $z_{1,\bfn}$, the 
$\psi_{1,\bfn}$, is unchanged and  equal to $(1,0)$.   

\begin{figure}[t]
\centerline{\epsfxsize=15cm\epsfbox{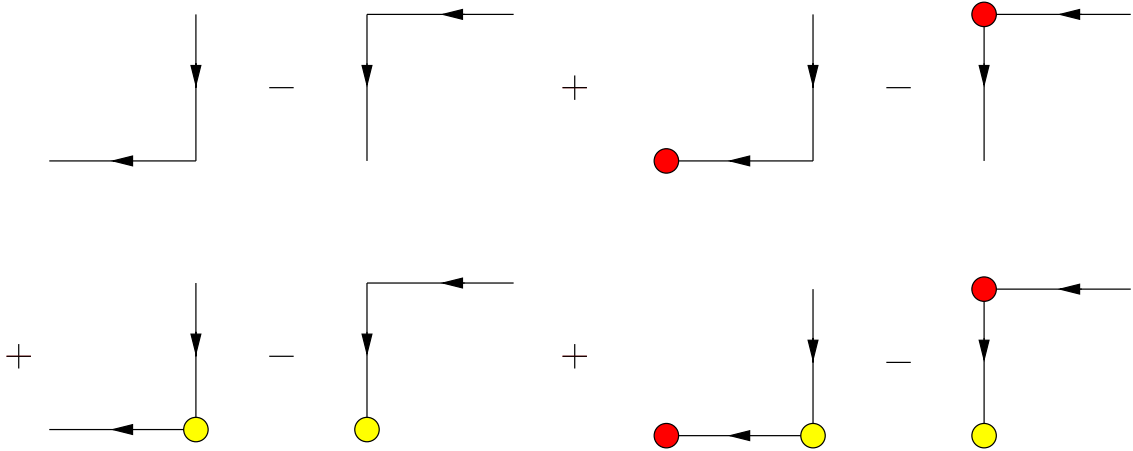}}
\smallskip
\caption{\sl The  $-\theta$ component of the fermi multiplet, 
  ${\bf \Xi}_{12,\bfn}$.  The vertical and horizontal  
arrows are unitary link fields, the red (yellow) circle is the $h_1$ $(h_2)$ 
scalar. The arrow starts at site 
$\bfn +{\bf \hat e_1}+{\bf \hat e_2} $ and  terminates at site $\bfn$. 
 The modulus square of the sum generates both Wilson action and part
of the scalar action.}
  \label{fig:fig1} 
\end{figure}

Next, we will show that expanding the action by using polar decomposition, 
 \Eq{polar} will  reproduce the Wilson action for the gauge fields 
and gauge invariant 
hopping terms for fermions  and scalars. Before we do that, let us rewrite 
the  superfields \Eq{superfield1} 
by using the new parametrization.  The superfield ${\bf
  \Lambda}_{\bfn}$ is same as  above and the others are  
\begin{equation}
\begin{aligned}
&{\bfZ}_{a,\bfn} = \frac{1}{\sqrt 2} (
\frac{1}{ a}U_{a,\bfn}+ h_{a,\bfn}U_{a,\bfn})  + 
\sqrt{2}\,\theta \,\psi_{a,\bfn}\ ,\\
 &&\\
&{\mybar z_{a,\bfn}} = \frac{1}{\sqrt 2} ( 
\frac{1}{a}U_{a,\bfn}^{\dagger}+ U_{a,\bfn}^{\dagger} h_{a,\bfn} ),   \\
 &&\\
&{\bf \Xi}_{12,\bfn}= \xi_{12,\bfn} -  \theta\,\,\Bigl[ \frac{1}{a^2}(
  U_{2,\bfn+ {\bf \hat e_1} }^{\dagger}   U_{1,\bfn}^{\dagger} - 
\,\,
   U_{1,\bfn+ {\bf \hat e_2} }^{\dagger}  U_{2,\bfn}^{\dagger})
 +  \frac{1}{a}(
  U_{2,\bfn+ {\bf \hat e_1} }^{\dagger}   U_{1,\bfn}^{\dagger}h_{1,\bfn} - 
\,\,
  U_{1,\bfn+ {\bf \hat e_2} }^{\dagger}h_{1,\bfn+ {\bf \hat e_2} }
  U_{2,\bfn}^{\dagger})+ 
\\ & \frac{1}{a}(
  U_{2,\bfn+ {\bf \hat e_1} }^{\dagger} h_{2,\bfn+ {\bf \hat e_1} }
  U_{1,\bfn}^{\dagger}
 - 
\,\,
  U_{1,\bfn+ {\bf \hat e_2} }^{\dagger}  U_{2,\bfn}^{\dagger} h_{2,\bfn})+ 
(
  U_{2,\bfn+ {\bf \hat e_1} }^{\dagger}h_{2,\bfn+ {\bf \hat e_1} }
  U_{1,\bfn}^{\dagger}
 h_{1,\bfn}- 
\,\,
   U_{1,\bfn+ {\bf \hat e_2} }^{\dagger}h_{1,\bfn+ {\bf \hat e_2} } 
 U_{2,\bfn}^{\dagger}h_{2,\bfn}) \Bigr]
 \ .
\end{aligned}
\label{eq:superfield2}
\end{equation}
The superfields  are covariant under gauge transformations and
the components transform homogeneously under all global symmetries. Also 
notice that the $\theta$  component of the fermi multiplet  
${\bf \Xi}_{\bfn}$ involves the square root of the standard Wilson action 
found in lattice QCD  and its decoration with scalars 
insertions. The $\theta$ component is depicted in Fig. \ref{fig:fig1}.
%
%
 
\subsection{The boson lattice   action}
{\it Gauge fields}

Our first goal is to show that our action involves the Wilson 
action for lattice gauge theory. The Wilson action is hidden in the third 
term of the bosonic action \Eq{ssact2b}. 
The $d_{\bfn}$ term does not contribute to the gauge field action. 
One easy geometric way to see that is,  the  $d_{\bfn}$ term  is coupled to 
composite fields which do not surround  an area, and  the Wilson action is 
a sum over trace of the elementary square plaquettes, (which can be
associated with the exponential of the Aharonov-Bohm flux in $U(1)$ 
target theories.)   

For our purpose, it is more useful to rewrite the third  term 
in the bosonic action \Eq{ssact2b} in a more suggestive form which makes 
the geometric visualization  easier: 
\begin{equation}
\begin{aligned}
S_{b2} = &\frac{2}{g^2} \sum_{\bfn} \tr \Bigl[
|( z_{1,\bfn} z_{2,\bfn
  +{\bf \hat e_1}} - z_{2,\bfn}  z_{1,\bfn + {\bf \hat e_2} })|^2    
\Bigr] \\
&=  \frac{2}{g^2} \sum_{\bfn} \tr \Bigl[ z_{1,\bfn} z_{2,\bfn
  +{\bf \hat e_1}} \mybar z_{1,\bfn+ {\bf \hat e_2}}  
\mybar z_{2,\bfn} 
- z_{1,\bfn} z_{2,\bfn
  +{\bf \hat e_1}} \mybar z_{2,\bfn+ {\bf \hat e_1}}  
\mybar z_{1,\bfn}  + h.c. \Bigr] 
\eqn{ssact2bG}
\end{aligned}
\end{equation} 
where $h.c.$ stands for hermitian conjugate.  
By plugging  the expansion \Eq{polar} into  above expression,  we obtain
(among other things which will be explained momentarily)
\begin{equation}
\begin{aligned}
S_{gauge}= &\frac{1}{2g^2a^4} \sum_{\bfn} \tr \Bigl[
|( U_{1,\bfn} U_{2,\bfn
  +{\bf \hat e_1}} - U_{2,\bfn}  U_{1,\bfn + {\bf \hat e_2} })|^2    
\Bigr] \\
&=  \frac{1}{2g^2a^4} \sum_{\bfn} \tr \Bigl[ U_{1,\bfn} U_{2,\bfn
  +{\bf \hat e_1}} U_{1,\bfn+ {\bf \hat e_2}}^{\dagger}  
U_{2,\bfn}^{\dagger} 
- 1  + h.c \Bigr]
\eqn{Wilson} \,,
\end{aligned}
\end{equation} 
This  is exactly the Wilson action for the pure lattice gauge theory. 
The   constant term in the action  sets  
the action to zero for the vacuum state. 
The  first term is a  square plaquette variable. We will refer to the 
constant  term   as  
``flipped L'' ($\usebox{\flippedL}$) since it is a product of unitary link
fields  
$U_{1,\bfn} U_{2,\bfn +{\bf \hat e_1}} U_{2,\bfn+ {\bf \hat e_1}}^{\dagger}  
U_{1,\bfn}^{\dagger}=1$ . Even though it seems  useless to
introduce such a name for identity, it will be a convenient 
tool  when we incorporate scalars.   
The scalars  enter as decorations on plaquette and flipped L terms.
Since the action \Eq{Wilson} is expressed in terms of the angular variables, 
the gauge fixing becomes  unnecessary.  \\
{\it Scalar-gauge and scalar-scalar interactions}

The scalar hopping terms and the the scalar potential term 
originate  from two sources. One is the third term in scalar action 
\Eq{ssact2b}, which also gives rise the Wilson action, 
and the other is the $d_{\bfn}$ term. We examine both in turn, starting 
with the second part \Eq{ssact2bG}. 

The second part of the action \Eq{ssact2bG} 
is the trace of the modulus square of the $\theta$ component 
of the fermi multiplet ${\bf \Xi}_{12,\bfn}$ (See Fig. \ref{fig:fig1}).
The first pair in Fig. \ref{fig:fig1} generates the field strength for the 
gauge field. The second pair  generates gauge covariant hopping 
terms for scalar $h_1$ in the ${\bf \hat e_2}$ direction. Similarly, the third 
pair   generates the hopping terms for the  scalar $h_2$ in the 
${\bf \hat e_1}$ direction. The last term give rise to the scalar-scalar 
interactions. In the continuum limit, these terms 
add up to $iv_{12} + D_2h_1 - D_1h_2 +[h_1, h_2] + O(a)$ where $v_{12}$ is the
gauge field  strength and $D_a= \partial_{a} + i[v_a, \cdot \ ]$ is the gauge
covariant derivative. 

Notice that this is not the most economical way 
to express the gauge covariant derivatives or scalar-scalar interactions 
on the lattice.  For example, consider the $h_2$  field. A more 
economical covariant derivative in direction one would be 
$h_{2, \bfn + {\bf \hat e_1}}- U_{1, \bfn}^{\dagger}h_{2, \bfn}U_{1,
  \bfn}$. However,  the exact supersymmetry dictates something  
different then that. Nevertheless, both turn out to yield the same 
gauge covariant 
derivative in the continuum, where the  difference between the two 
is  suppressed by lattice spacing. 
Also,  the scalar-scalar interaction  term $[h_1, h_2]^2$ in the action  
does not follow from a local potential 
on the lattice. The way it arises seems  a bit extravagant. 
For example, it emerges from a plaquette and  flipped L decorated 
with four scalars at various sites. This is slightly nonlocal, unlike 
our usual way of writing the scalar interactions.
 But similar to the covariant derivative, the difference of
the slightly nonlocal potential and local potential is suppressed with the 
lattice spacing toward the continuum limit. 
  
The full expression for the \Eq{ssact2bG}, expressed in terms of
unitary link fields  and hermitian scalars, 
is the trace of the modulus square of the sum of figures
presented  in Fig. \ref{fig:fig1}.  The square   involves both the 
plaquettes 
(denoted as $\Yfund$) and the flipped L's (denoted as $\usebox{\flippedL}$) 
 with scalar decorations. Hence, it is more convenient to 
assemble \Eq{ssact2bG}  into a simple (symbolic) expression given as: 
\begin{equation}
\begin{aligned}
S_{b2} =  \sum_{k=0}^{4} \frac{1}{2a^{4-k}g^2} \sum_{( \ \Yfund, \  \usebox{
\flippedL})}
 \tr \Bigl[
(U[
\, \Yfund \,,k ]
-U[\usebox{\flippedL}\!, k] ) 
  + h.c. \Bigr] 
\eqn{Wilson2G} \,, \\
\end{aligned}
\end{equation} 
where $k$ is the number of the scalar insertions onto  
the plaquette  or the flipped L. 
The rule for the scalar insertions was indeed fixed when we chose the left
decomposition in polar decomposition \Eq{polar}. The scalar 
 $h_{a, \bfn}$ can only be  inserted at the
starting point of  $U_{a, \bfn}$ and at the end  point of 
$U_{a,  \bfn}^{\dagger}$. 
  
The $k=0$  term is the Wilson action as discussed. There are
harmless  $k=1$ and $k=3$ terms, the decorations with one and three scalars 
respectively. Examination of these terms shows that due to exact
cancellations  among the various components in the sum, the
lowest dimensional operators arising from these terms are dimension five 
(in the normalization used in  \cite{Cohen:2003aa}), 
 suppressed by lattice spacing, and hence irrelevant.  
The $k=2$ terms in the sum 
provide gauge
invariant hopping terms for scalars, namely $h_1$  ($h_2$)   hopping in the  
${\bf \hat e_2}$ (${\bf \hat e_1}$) direction. There are two types of 
 Lorentz symmetry  violating cross term, which cancels by the 
contribution coming from the $d_{\bfn}$ term.  
The $k=4$ sector generates  the quartic   scalar  interaction of the
continuum theory.  
 
The auxiliary $d_{\bfn}$ field enters into the action quadratically and 
can be integrated out.  The equation of motion gives 
\begin{equation}
\begin{aligned}
 d_\bfn  & =  -i \sum_{a} \left(\mybar z_{a,\bfn-\ah}  
    z_{a,\bfn-\ah} - z_{a,\bfn}  \mybar
    z_{a,\bfn}\right) \\
& = -i \sum_a \left( \frac{1}{a}(U_{a,\bfn-\ah}^{\dagger} h_{a,\bfn-\ah}U_{a,\bfn-\ah} -  h_{a,\bfn})
+ \half (U_{a,\bfn-\ah}^{\dagger} h_{a,\bfn-\ah}^2 U_{a,\bfn-\ah} - 
 h_{a,\bfn}^2) \right)
\end{aligned}
\label{eq:dterm}
\end{equation} 
which generates the 
gauge covariant hopping  term for scalars. In the continuum, this becomes 
$D_1h_1 +D_2h_2 +  O(a)$. 
 Substituting the \Eq{dterm} 
into the  first part of the bosonic 
action, we obtain   
\begin{equation}
\begin{aligned}
S_{b1} = 
&=  \frac{1}{2g^2} \sum_{\bfn} \tr \Bigl[  
 (\sum_a  \frac{1}{a}(U_{a,\bfn-\ah}^{\dagger} h_{a,\bfn-\ah}U_{a,\bfn-\ah} -  
h_{a,\bfn}))^2 + \ldots
\Bigr]
\eqn{ssact2b2}
\end{aligned}
\end{equation} 
where the ellipsis indicates the terms which becomes irrelevant in  the 
infrared . The continuum limit of  the action \Eq{ssact2b2}, 
generates 
hopping terms for $h_a$ in  ${\bf \hat e_a}$ directions and a mixed Lorentz
symmetry violating term. The latter cancels exactly with the cross 
term arising from the \Eq{Wilson2G}. Upon adding these two parts,
$S_{b1}+S_{b2}$,  we 
obtain  the bosonic part of the target theory action, the $\CN=(2,2)$ SYM 
theory in $d=2$ dimensions. Namely, 
\begin{equation}
\begin{aligned}
S_{b}  
&=  \frac{1}{2g_2^2} \int \ d^2x   \ \tr \Bigl[ (D_1 h_1 + D_2 h_2)^2 +     
| (D_1 h_2 - D_2 h_1) +i (v_{12} - i[h_1,h_2])|^2 \\
& = \frac{1}{g_2^2} \int \ d^2x   \ \tr \Bigl[ \ \frac{1}{4} v_{\mu\nu}^2 + 
\frac{1}{2} 
(D_{\mu}h_a)^2 - \frac{1}{4} [h_a, h_b]^2
\Bigr]
\eqn{act2b} 
\end{aligned}
\end{equation} 
where $\mu, \nu=1,2 $ and $a,b=1,2 $.
The three cross terms add up to a Lorentz invariant 
surface term proportional to 
$$\epsilon^{\mu\nu} \left( D_{\mu}h_1 D_{\nu}h_2 -
\frac{i}{2} v_{\mu \nu} [h_1,h_2] \right),
$$   
which can be shown to 
be the dimensional reduction of the topological term, 
$\tr \vec E. \vec B $ from $d=4$ to $d=2$ dimensions.   

\subsection{The fermion  lattice action}
The fermion gauge field interactions and the hopping terms  are particularly
simple. Substituting \Eq{polar} into fermion action \Eq{ssact2f}, we find 
the fermion-gauge field and the fermion-scalar interaction
terms. The former is   
\begin{equation}
\begin{aligned}
S_{f-g} = \frac{1}{a g^2} \sum_{\bfn} \tr \Bigl[
 \lambda_\bfn (
    U_{a,\bfn-\ah}^{\dagger}  
    \psi_{a,\bfn-\ah} - \psi_{a,\bfn}  
    U_{a,\bfn}^{\dagger} )
&
+ \xi_{ab,\bfn}  (  U_{a,\bfn}  \psi_{b,\bfn
  +{\bf \hat e_a}} - \psi_{b,\bfn}  U_{a,\bfn + {\bf \hat e_b} } )
\Bigr]
\eqn{fermionhopping} \,,
\end{aligned}
\end{equation} 
These hopping terms can be classified in two types. 
%
\begin{figure}[t]
\centerline{\epsfxsize=15cm\epsfbox{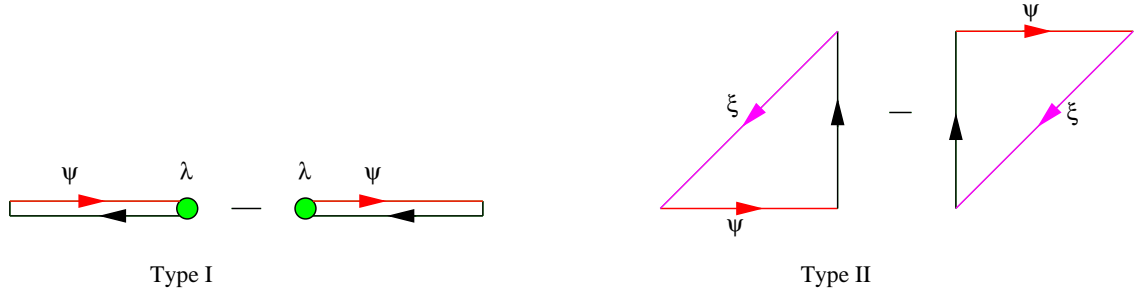}}
\smallskip
\caption{\sl Representatives of type I and type II fermionic hopping terms. 
The black lines are unitary link fields. The green circle represents a site 
fermion $\lambda_{\bfn}$. Red lines are link fermions, $\psi_{a, \bfn}$. The 
diagonal of the triangular  plaquette is the diagonal link fermion
$\xi_{\bfn}$
starting at  $\bfn +{\bf \hat e_1}+{\bf \hat e_2} $ and  terminating  
at site $\bfn$.}
  \label{fig:fig2} 
\end{figure}
%
The first type is the usual 
hopping term, which does not surround an area, and the second type makes a
triangular plaquette. Both are  manifestly gauge invariant. 
In the  type I case in Fig. \ref{fig:fig2}, 
the unitary link fields  properly  parallel transport 
the link fermions $\psi_{a,\bfn}$ and  $\psi_{a,\bfn - \ah}$ to the site
$\bfn$,
by a 
backward and forward parallel transport, respectively.
The signed sum of these two terms  couples 
to the site fermion  $\lambda_\bfn$. 
In the type II case,   $\xi_{ab,\bfn}$ resides 
on the diagonal of the unit cell. The  
gauge invariant hopping term is a signed sum of two triangular plaquettes, 
whose two sides are fermionic and one side is the unitary link field.  

Unlike the traditional hopping terms for adjoint fermions, 
which typically involves two parallel transports to make a gauge invariant 
hopping term, such as 
$\Tr (\chi_{\bfn}U_{\bfn,a}\chi_{\bfn + \ah}  U_{\bfn,a}^{\dagger})$, our hopping
terms only involve one unitary link field. 
The difference comes about because 
the fermions in our formulation are scattered to both sites and
links, and it suffices to have a single unitary link field to build a 
gauge invariant hopping term. However, 
for example, with  adjoint fermions, the fermions 
are all residing on the  sites and the simplest gauge invariant object
requires two parallel transports. 

The fermion scalar interaction can easily be incorporated  
from \Eq{ssact2f}.  We obtain 
\begin{equation}
\begin{aligned}
S_{f-s} =& \frac{1}{ g^2} \sum_{\bfn} \tr \Bigl[
 \lambda_\bfn (
    U_{a,\bfn-\ah}^{\dagger}   h_{a,\bfn-\ah}
    \psi_{a,\bfn-\ah} - \psi_{a,\bfn}  
    U_{a,\bfn}^{\dagger} h_{a,\bfn} ) \\
&
+ \xi_{ab,\bfn}  (  h_{a,\bfn} U_{a,\bfn}  \psi_{b,\bfn
  +{\bf \hat e_a}} - \psi_{b,\bfn}  h_{a,\bfn + {\bf \hat e_b} } 
 U_{a,\bfn + {\bf \hat e_b} } )
\Bigr]
\eqn{fermionscalar} \,,
\end{aligned}
\end{equation} 
The fermion-scalar interaction part of the action can be  regarded 
as decoration of the fermion hopping terms with scalars. The unitary link
fields are needed to make gauge invariant fermion-scalar interaction 
terms since most of the  fermions are residing on the links. 

\section{Observables and  global symmetries}
 \label{sec:observables}

In pure gauge theories, the observables  are the the gauge invariant 
Wilson loops,  constructed out of  an ordered product of the 
unitary link fields along the loop. 
In  supersymmetric lattice gauge theories (constructed within the approach of 
this work), the link fields can be both 
fermions and unitary matrices.(Scalar are placed at the sites by the 
polar decomposition \Eq{polar}.) Thus, the natural generalization of the 
Wilson loops are  the 
gauge invariant  loops, constructed out of an ordered product of group valued 
unitary 
 and algebra valued fermionic  link fields, 
also possibly decorated with  arbitrary 
insertions  of site fermions and scalars along the loop \cite{Kovtun:2003hr}.
Obviously, a subset of observables are the usual Wilson loops. Bosonic and
fermionic loops
 solely composed of fermions are also possible, such as  
$\Tr[\psi_{1,\bfn}\psi_{2,\bfn +  {\bf \hat e_1}  }\xi_{12,\bfn}]$, a
fermionic  triangular plaquette.  The correlation function of such loops 
are also among observables. 

The questions about the nature of these loops, such as  
``how do we classify them'', or ``what do these loops corresponds to'' 
can be  partially  
be answered by analyzing the global symmetries. For example, 
in pure gauge theories Wilson loops are neutral under the center of the gauge 
group and Polyakov loops transform in the one dimensional representation 
of the center. For the $U(N)$ lattice gauge theory in $d$ dimensions, the 
action possesses  a
 $U(1)^d$ symmetry  associated with the invariance 
of the action under $U_{a,\bfn}\rightarrow e^{i \alpha_b \delta_{ab}} 
U_{a,\bfn}$,  an independent $U(1)$  rotations in each direction\footnote{
More  precisely, consider a gauge theory with a gauge group $G$ 
on a $d$ dimensional torus. Let  $C(G)$ denote the center of the gauge group 
$G$. Then, the Polyakov loops are charged under   $C(G)^d$. 
In particular, for $U(N)$ gauge group, the center is  $U(1)$ and  
the Polyakov loops are
charged under $U(1)^d$.}. 
A particular $U(1)$ charge of a Polyakov loop can be characterized by  the 
 the number of winding of the loop in that direction. 
\setlength{\extrarowheight}{5pt}
\begin{table}
\centerline{
\begin{tabular}
{|c||c|c|c|c|c|c||c|c|c|c|}
\hline
&$  U_{1,\bfn}  $&$  U_{1, \bfn}^{\dagger} $&$ U_{2,\bfn}$&$\mybar
U_{2,\bfn}^{\dagger} $ &$  h_{1,\bfn}  $ &$  h_{2,\bfn}  $
&$ \lambda_{\bfn} $&$
{ \xi}_{12,\bfn} $&$\psi_{1, \bfn}$&$\psi_{2,\bfn}$
\\ \hline
$ r_1 $&$ +1 $&$ -1 $&$ \,\ 0 $&$ \,\ 0 $&$ \,\ 0 $&$ \,\ 0 $&
  $ \,\ 0 $&$ -1 $&$  1 $&$  0 $
\\
$ r_2 $&$\,\ 0 $&$ \,\ 0 $&$ +1 $&$ -1 $& $ \,\ 0 $&$ \,\ 0 $& 
$ \,\ 0 $&$ -1 $&$ \,\ 0 $&$ \ 1 $  
\\
%
$ Y $ &$  \,\ 0 $&$ \,\ 0 $&$ \,\ 0 $&$ \,\ 0 $& $ \,\ 0 $&$ \,\ 0 $
&$ +\half $&$ +\half
$&$ -\half $&$ -\half $
\\ \hline
 \end{tabular}
}
\caption{\sl The  $U(1)^3$  symmetry charges 
 of the  lattice theory. The first two are associated with the 
center symmetry of the two dimensional lattice gauge theory and the $Y$ is 
R-symmetry. 
\label{tab:tab3}}
\end{table}

The supersymmetric lattice action that we examined in section 
\ref{sec:appl-comp-latt} has a $U(1)^3$ symmetry.  The charges under the 
global 
 $U(1)^3$  symmetry are listed in Table \ref{tab:tab3}. 
(This table has the same content as 
 Table \ref{tab:tab2}, but it has been rewritten in terms of new variables 
for convenience.)  
The $U(1)^2$ associated with $(r_1, r_2)$ are  
associated with the square of the center symmetry of the gauge group 
$U(N)$ in the   two dimensional lattice. The last  $U(1)$ was clear 
from the outset; it is the global R-symmetry. 

Let us now consider the observables. Assume the lattice is compact in all
directions, a discretized torus. 
Let us first consider the loops which do not wind around the lattice.
 All the closed loops of this type are neutral under 
$U(1)^2$, associated with center symmetry. This can be easily read of from
the Table \ref{tab:tab3}.    
 However, it is not hard to see that only a subclass of observables are
neutral  under $U(1)_R$. This is the case if the total 
number of $\lambda_{\bfn}$, and $\xi_{12,\bfn}$ (whose R charges are $\half$) 
 is exactly same as the total number of $\psi_{1,\bfn}$, and  
$\psi_{2,\bfn}$ (whose R charges are $-\half$).
The observables with a nonvanishing R-charge and transforming homogeneously 
under $U(1)_R$ rotation have zero expectation value. (As long as the 
$U(1)_R$ is not spontaneously broken.) This type of observable can 
be used to probe  the spontaneous breaking of $U(1)_R$ symmetry. 
(see, for example, the review 
\cite{Montvay:2001aj} and  
\cite{Fleming:2000fa} about the spontaneous breaking of R-symmetry, $Z_{2N}$, 
in $\CN=1$ $SU(N)$  SYM theory  on the four dimensional lattice.)  
In particular, the expectation 
value of any observable with an odd number of fermions is zero, which is
fermion number conservation modulo two. 
 
There are also observables, the 
counterparts of the Polyakov loops, which are
charged under the center $U(1)^2$ and possibly under $U(1)_R$. In general, 
an observable is associated with a  charge triplet, 
$(n_1, n_2, \frac{n_R}{2})$   under $U(1)^3$. The  $n_1$ $(n_2)$ is 
the net number of  windings in direction one (two).  However, there is 
subtlety here, which does not show up in the pure gauge theory. In pure gauge
theory,   a Polyakov loop in a direction can be undone by traversing 
the same loop in the 
opposite direction, since the constituents are unitary matrices. 
 In our case,    
since some of the link fields are algebra valued fermions, the backtracking 
(Polyakov) loops do not necessarily cancel. 
The integer $n_R$  is the net R-charge of the corresponding loop.  
Obviously, an observable which is charged under 
any one of the global  symmetries has vanishing expectation value as long 
as the global symmetry associated with the given charge 
is  not spontaneously broken. 

\section{Prospects} 
The technique we have described here can be applied to all the
Euclidean lattice constructions \cite{Cohen:2003aa, Cohen:2003qw, 
Kaplan:2005ta} for  extended supersymmetric gauge theories listed 
in Table \ref{tab:tab1}. It is also possible to work with spatial 
lattices, \cite{Kaplan:2002wv} 
which are more suitable for Hamiltonian formulation of 
 supersymmetric lattice gauge theories.
 
It would be useful  to investigate 
 the strong coupling expansion for the observables 
(loops and  the connected 
correlators of loops) in the large-$N_c$ limit, where $N_c$ is the number 
of colors. 
Or equivalently, one can examine the  lattice regularized loop 
equations for the supersymmetric lattice gauge theories by generalizing  
\cite{Kovtun:2003hr} to the theories with massless fermions. It would be 
interesting to examine to what extend the large $N_c$ reduction holds in
the supersymmetric gauge theories 
\cite{Narayanan:2003fc, Neuberger:2002bk} and understand 
their large-$N_c$ phase diagram. 
 
\acknowledgments
I am  grateful for numerous conversations about this work with Andrew Cohen, 
David B. Kaplan and Larry Yaffe. I also  wish to thank Larry Yaffe for raising
the questions about the compact formulation.  
I thank Ron Babich for reading the manuscript and suggestions.
This work was supported by DOE grant DE-FG02-91ER40676.

\bibliography{susycompact}
\bibliographystyle{JHEP} 
\end{document}